# ChatLearn: Leveraging AI to Transform Non-Native Speaker Communication Challenges as Language Learning Opportunities

Peinuan Qin
School of Computing, National
University of Singapore
Singapore, Singapore
e1322754@u.nus.edu

Yugin Tan
School of Computing, National
University of Singapore
Singapore, Singapore
tan.yugin@u.nus.edu

Jingzhu Chen
School of Computer Science and
Technology, Tongji University
Shanghai, China
2253543@tongji.edu.cn

Nattapat Boonprakong
School of Computing, National
University of Singapore
Singapore, Singapore
nattapat.boonprakong@nus.edu.sg

Zicheng Zhu
School of Computing, National
University of Singapore
Singapore, Singapore
zicheng@u.nus.edu

Naomi Yamashita
Graduate School of Informatics, Kyoto
University
Kyoto, Japan

Yi-Chieh Lee
National University of Singapore
Singapore, Singapore
yclee@nus.edu.sg

## Abstract

Non-native speakers (NNSs) face significant language barriers in multilingual communication with native speakers (NSs). While AI-mediated communication (AIMC) tools offer efficient one-time assistance, they often overlook opportunities for NNSs' continuous language acquisition. We introduce ChatLearn, an enhanced AIMC system that leverages NNSs' communication difficulties as learning opportunities. Beyond comprehension and expression assistance, ChatLearn simultaneously captures NNSs' language challenges, and subsequently provides them with spaced review as the conversation progresses. We conducted a mixed-methods study using a communication task with 43 NNS-NS pairs, after which ChatLearn NNSs recalled significantly more expressions than the baseline group, while there was no substantial decline in communication experience. Our findings highlight the value of contextual learning in NNS-NS communication, providing a new direction for AIMC systems that foster both immediate collaboration and continuous language development.

## CCS Concepts

• **Human-centered computing → Empirical studies in HCI**; **Laboratory experiments**; • **Applied computing → Education**.

## Keywords

AI-Mediated Communication; Language Acquisition; Communication Support

## 1 Introduction

Non-native speakers (NNSs) often struggle to communicate with native speakers (NSs) due to limited vocabulary [41, 85], poor grammar [13], and sentence structure difficulties [20]. These obstacles can hinder effective collaboration and lead to suboptimal teamwork

[31, 81]. AI-mediated communication (AIMC) approaches [30, 53] utilize AI assistance to enhance communication processes, and have proven effective in overcoming language barriers [1, 21, 45]. This creates practical benefits such as reduced social discomfort [22] and improved logical structure of messages [64]. However, existing AIMC methods are primarily designed to offer NNSs one-time assistance, while neglecting support for their continuous language acquisition. As a result, they enhance NNSs' immediate performance without corresponding skill development [38, 51].

Regarding this issue, previous theories on second language (L2) learning have indicated that communication between NNSs and NSs can serve as an opportunity for learning. The *input hypothesis* [42] describes content slightly beyond one's current ability - such as unfamiliar, second-language expressions appearing in conversation - as "desirable difficulties" [5, 40]. These are appropriately challenging inputs that are still comprehensible, encouraging learning and expanding NNSs' capacity boundaries [86]. The *output hypothesis* [79], meanwhile, emphasizes that the process of language production helps learners notice gaps in their knowledge and learn as a result of trying to fill those gaps. These studies suggest that **NNSs' language difficulties with comprehension and expression [7, 94] in communication, can be considered ideal learning opportunities**. From the perspective of contextual learning, conversations provide relevant, socially situated tasks where unfamiliar grammar and vocabulary are embedded in real-world use [6, 14, 26, 75, 82], allowing for effective learning [82]. **The authentic and rich context of NNS-NS dialogue can provide the necessary environment for contextual learning to occur.**

However, integrating learning into real-time communication presents potential risks. NNSs already experience cognitive burden when conversing with NSs [7, 28, 76]. Embedding additional learning tasks may exacerbate cognitive overload [54], interrupt conversational flow [52, 64], and impair the overall communication experience [48]. Thus, the challenge lies in designing AIMC systems that strike a balance between effective real-time communication and language learning. To minimize disruption and avoid cognitive overload, prior research suggested that effective language acquisition can be facilitated through *incidental exposure* [10, 23, 33], emphasizing that learning occurs through multiple unconscious repetitions.

Based on these learning and design principles, we introduce ChatLearn, an AIMC system that supports continuous language learning for NNSs in the setting of online text-based communication. Beyond comprehension and expression support through translation, ChatLearn offers additional features that transform language challenges into learning opportunities. NNS users can manually select challenging expressions they encounter in conversation for explanation; they can also use their first language (L1) as a scaffold to construct messages when language gaps hinder their expression. These instances where NNSs require language support are captured by ChatLearn as potential learning materials, which are subsequently retrieved and displayed when relevant in subsequent conversation to provide incidental learning opportunities [23, 33].

To investigate the impact of ChatLearn on L2 learning and NNS-NS communication experience, in this paper, we ask the following questions.

**RQ1:** How does ChatLearn influence NNSs' language acquisition?
  **RQ1.1:** How does ChatLearn influence NNSs' recall of unfamiliar L2 expressions compared to a communication-oriented baseline?
  **RQ1.2:** What are the mechanism(s) in ChatLearn that affect NNSs' recall performance?
**RQ2:** What is the impact of ChatLearn's learning-oriented features on NNSs' communication behaviors and perceived communication experience, compared to the communication-oriented baseline?
**RQ3:** How do NNSs use ChatLearn's features while communicating, and how does the theoretical framework behind its components operate to support the learning process?

We conducted a mixed-methods, between-subjects experiment in which 43 pairs of native and non-native speakers conversed using an AIMC system. We compared two conditions: a communication-oriented baseline (hereafter referred to as **Baseline**) and **ChatLearn**. The *Baseline* condition (21 NNS-NS pairs) simulated existing AIMC tools that are designed solely to overcome communication barriers. In this condition, NNS participants were only provided with translations for comprehension and expression support (see Section 3.1.1). In contrast, the *ChatLearn* condition (22 NNS-NS pairs) incorporated both the communication support of the Baseline and additional learning-oriented features (see Section 3.1.2). To explore the impact of ChatLearn in an authentic communication setting, participants were not given any explicit learning requirements and were instructed simply to chat with their partners.

We found that NNSs who used *ChatLearn* significantly recalled more expressions in a free-recall post-test. Mediation analysis indicates that ChatLearn's learning features *directly* affect recall performance, while NNSs' learning motivation *indirectly* intervenes in recall performance by influencing germane cognitive load. While quantitative data and NNSs acknowledged that ChatLearn did not significantly affect the overall communication experience, their perceived responsiveness and clarity in self-expression were lower than that of *Baseline* participants. Further analysis indicates that this may not be a negative signal but rather that ChatLearn has established a new balance between the learning process and communication goals. Our results also reveal the process of NNSs interacting with ChatLearn components and highlight which key interactions and experiences are fundamental to the effectiveness of ChatLearn. These findings can be used to guide the design of future AIMC systems. In summary, this paper offers three key contributions to HCI research:

(1) We present ChatLearn, an AIMC system that not only supports communication between NNS and NS users, but also transforms communication challenges into learning opportunities.
(2) We provide empirical evidence from a mixed-method study with 43 NNS-NS pairs, demonstrating that ChatLearn improves expression recall while maintaining acceptable communication experiences.
(3) We contribute design implications outlining how AIMC systems can be designed to balance communication effectiveness and language learning, and how learning opportunities can be extended beyond single interactions.

## 2 Related Work

### 2.1 AI-Mediated Communication Tools for Non-Native Speakers

AI-Mediated Communication (AIMC) systems provide automated tools for modifying, augmenting, or generating messages to aid communication goals [30, 53]. These systems have demonstrated effectiveness in myriad contexts, including email writing [12, 46, 49], multilingual negotiation in economic games [47], and multilingual collaborative problem solving [64]. AIMC systems are particularly helpful for NNSs communicating in an unfamiliar language. NNS often encounter comprehension and expression difficulties when communicating with NSs, hindering their understanding and opportunities to express themselves [7, 76, 94]. AIMC tools address these issues by alleviating cognitive loads [74], improving logical cohesion of constructed messages [64], and reducing the social discomfort of communicating with an NS [22] while improving conversation performance [47]. For instance, Lim and Yang [47] showed that providing translation support in multilingual negotiation scenarios enhanced communication and allowed NNSs to achieve more equal negotiation outcomes. Meanwhile, Wang et al. [87] integrated AI translation into instant messages to improve NNSs' idea generation .

While such systems effectively alleviate immediate difficulties that NNSs may encounter, this comes at the cost of enhancing NNSs' overall language abilities. Ideally, the system should reduce extraneous cognitive load [61]—the unnecessary mental effort caused by interface complexity or task-irrelevant processing—while maintaining germane cognitive load [17], the productive effort users invest in processing, reflecting on, and internalizing linguistic knowledge. However, in prioritizing solving communication problems, many AIMC systems overly offload cognitive engagement [24, 68, 74], leading to weakened cognitive processing [15] and critical thinking [73] about the language challenges that NNSs encounter. This is a common trend with AI systems that assist with complex tasks by bypassing the effort necessary to achieve a learning objective [68], and can eventually lead to over-dependence on AI tools [32, 57] and stagnating language ability [51, 83]. Lee [43] further pointed out that translation tools only aid language acquisition when used in combination with teaching strategies. This exposes a gap in the design of current communication support tools for NNSs: while effective in the short term, they lack sufficient support for continuous language development [51]. To bridge this gap, we aim to design a novel AIMC system that provides communication support while integrating language learning.

## 2.2 Foundations of Language Acquisition in NNS-NS Communication

Our proposed AIMC design is grounded in a number of well-established language-learning theories.

*2.2.1 Mechanisms for Effective Language Acquisition.* The *input hypothesis*, proposed by linguist Krashen [42], suggests that language acquisition occurs most effectively when users receive comprehensible input (e.g. messages) just beyond one's current proficiency level, which helps the learner naturally acquire the next steps of the language in a subconscious manner. In contrast, overly simple material fails to adequately challenge the user, while overly difficult material instead causes frustration. This is linked to the broader concept of *desirable difficulties* that enhance long-term retention of knowledge [5, 40]; such methods of learning challenge the learner in a productive way, leading to more effective learning than easier but less effective alternatives such as re-reading or blocked practice [16]. The *output hypothesis* [78, 79], proposed by Swain [79], meanwhile concerns producing instead of receiving language, asserting that such activities as speaking or writing are a crucial part of second language learning. Through attempts to produce language, learners notice "gaps" in their knowledge, test hypotheses about the language, and internalize grammar and vocabulary, ultimately leading to improved accuracy and fluency. Related to both inputs and outputs, the *noticing hypothesis* [71] suggests that learners must consciously identify and pay attention to specific linguistic features, in order to better learn and internalize them.

Collectively, these emphasize the potential of appropriately difficult challenges in both reading and writing for encouraging language acquisition, forming a basis for systems that leverage comprehension [7] and expression [76] challenges to provide effective learning opportunities. Therefore, we proposed the first design goal: [G1] Help NNSs notice and transform their language challenges into learning opportunities.

*2.2.2 Contextual and Incidental Learning.* Contextual learning is grounded in the idea that *"[t]he activity in which knowledge is developed... is an integral part of what is learned"* [6, 93]. Such learning happens when an individual actively learns something while in a particular situation in which it happens to be relevant. In language learning, this translates to exploiting real-world scenarios in which a user might need to use or understand phrases or expressions in an unfamiliar language, and using these as valuable learning opportunities. Contextual learning has several advantages [19]. In contrast to traditional textbook learning, where students receive instruction in predetermined settings or time slots, contextual learning provides information in a "just-in-time" manner, delivering it as and when it is needed and relevant [67]. Unlike flashcards [59] or other classroom learning methods, contextual learning involves exposure to linguistic expressions in the specific context in which they occur. This increases recall of such expressions the next time the learner is in the same context, such as a conversation about a particular topic - a phenomenon known as encoding specificity [82].

*Incidental learning*, meanwhile, is defined as *"a natural byproduct that arises during the cognitive activities of performing understanding tasks such as reading [or] listening"* [23]. This is also highly relevant to language learning, as research shows that second language learners generally acquire a significant portion of their vocabulary through such incidental exposure [65, 89]. Two factors are particularly relevant here. Firstly, both frequency and manner of repetition of words are crucial to incidental learning. Increased encounters of a given expression, while the concept of *spaced repetition* [39] suggest that these encounters need to occur gradually and over a period of time - rather than being crammed into a short duration - for better long-term memory [10, 60]. Secondly, the relevance of a given expression and its surrounding context to a learner are also critical. *Language socialization theory* builds on this, positing that individuals primarily learn language not through explicit instruction, but implicitly when they participate in social and cultural interactions [56]: active participation in such activities is a strong driver of acquisition of the language used.

These studies emphasize the importance of language learning approaches that should be as interactive, contextualized, and distributed as possible, suggesting that the real, interactive environment of NNS-NS communication is suitable for developing a learning experience. Therefore, we proposed the second design goal: [G2] Embed incidental learning in an authentic context.

## 2.3 Learning Through Conversation

This study explores conversations between native and non-native speakers as a highly valuable resource for language acquisition, in particular through contextual learning. Conversations inherently provide multiple advantages on this front. Firstly, unfamiliar expressions may appear in the natural flow of conversation, providing a situated context in which a NNS might want to understand and subsequently use that expression. This enhances learning potential, in accordance with the idea of encoding specificity [26] mentioned previously. Secondly, sustained conversations often contain multiple instances of similar phrases or meanings appearing over the duration of the conversation. This gives NNSs periodic and repeated exposure to particular expressions that they may be unfamiliar with, providing multiple opportunities for learning. This concept is known as spaced repetition [39], shown to be highly effective for language acquisition [72]; it also adheres to the gradual and longitudinal nature of contextual learning [27].

This context further aligns with other language learning theories mentioned above. When appropriately situated, NS-NNS conversations can provide the sort of desirable difficulties [5] that align with the input hypothesis [42]: for example, when NSs use expressions that are unfamiliar to the NNS, situated within the context of an otherwise comprehensible message. The need for NNSs to reply in a non-native language also aligns with the output hypothesis [79] and its emphasis on producing utterances in that language.

However, there are also potential challenges with learning through conversations. Systems that encourage learning in the midst of conversation may exacerbate existing difficulties with cognitive load [54], leading to worse communication efficiency [52, 64] and experiences [48] without achieving meaningful learning. Therefore, we proposed the third design goal: [G3] Balance communication and learning with lightweight learning support.

In this study, we investigate how these opportunities can be exploited through targeted learning features, while also investigating the potential negative effects of such a system.

## 3 Method

We operationalized our study in a context in which English is the target medium of communication, while Chinese native speakers who are not fluent in English are required to communicate with English native speakers.

To explore the opportunities and challenges with AIMC approaches that simultaneously support communication and learning objectives during NNS-NS dialogue, we designed two experimental conditions. NNS Participants in communication-oriented baseline condition used an interface with comprehension and expression support (see Section 3.1.1) to only facilitate their communication with NSs. For convenience, this condition refers to *Baseline* hereafter. In contrast, ***ChatLearn*** participants were additionally provided with learning acquisition features (Section 3.1.2).

In each condition, participants were paired on NNS-NS dyads and engaged in a text-based communication task, with NNSs having the baseline system or ChatLearn to help them. We then evaluated the difference between the two conditions on communication experience, behaviors, and language acquisition.

### 3.1 System Design and Implementation

We describe these features **from the perspective of the native-Chinese (the NNS participants)**, as the native-English (the NS participants) only acted as confederates and only interacted with a standard online-messaging system. For clarity and brevity, **"Chinese" refers to the NNSs' first language and "English" to their target language**. We first describe the features common across both conditions, before elaborating on the additional learning features introduced by *ChatLearn*. To clearly compare the functional differences between *Baseline* and *ChatLearn*, we used the Table 1 to list their respective components. Both the *Baseline* and *ChatLearn* interfaces were built based on the WeChat[1] interface. Since WeChat does not provide open interfaces to support secondary development by researchers, we used vue.js[2] and django[3] as the frontend and backend framework to develop the basic communication interface and built the corresponding functions of *Baseline* and *ChatLearn* on this basis.

#### 3.1.1 Communication-Oriented Baseline.
The *Baseline* system was designed to provide communication support similar to existing

---

[1]https://www.wechat.com/
[2]https://vuejs.org/
[3]https://www.djangoproject.com/

AIMC systems [1, 21, 45, 47], without specifically aiming to support learning. *Baseline* addresses known communication difficulties that NNSs experience, specifically expression construction [76] and comprehension [7]. Expression construction was supported by allowing NNSs to write messages partially or wholly in Chinese, then click a button to translate them into English (see (3) in Fig. 1). Expression comprehension was in the form of allowing NNSs to click a button under each received English message, and translate it directly into Chinese (see (2) in Fig. 1). These were integrated within a standard chat window (see (1) in Fig. 1). In both (2) and (3), for more contextually-accurate translations, we included the 6 most recent turns of chat history [63, 77] in the AI prompt (see Table 4).

#### 3.1.2 ChatLearn.
Different from *Baseline*, which focuses solely on providing *communication* assistance, ChatLearn is implemented with additional learning features on top of the *Baseline* feature set. Guided by established theories of second language acquisition (Section 2), our design of ChatLearn aims to achieve three main goals:

**G1:** Help NNSs notice and transform their language challenges into learning opportunities.

**G2:** Embed incidental learning in an authentic context.

**G3:** Balance communication and learning with lightweight learning support.

Below, we detail the rationale behind each goal, along with corresponding design features.

***G1: Noticing Language Challenges as Learning Opportunities.*** Grounded in the noticing hypothesis [71], we aimed to encourage NNSs to actively notice challenging expressions in both their inputs [42] and outputs [79], as a means of achieving more effective learning (see Section 2.2.1). Thus, we designed two components, the *(4) Expression Explorer* and the *(5) Expression Extractor*, to transform difficulties in comprehension and expression construction, respectively, into learning opportunities. Note that, here, *"expression"* may refer to individual words, phrases, or entire sentences.

When users encounter comprehension challenges, they can use Expression Explorer by selecting the specific expressions in received messages to obtain precise explanations and examples to contextualize their usage (see Table 5 for the prompt template). This encourages users to proactively identify challenging expressions for active learning [71]. The explanation is presented in the NNSs' first language for understanding, while the subsequent example is shown in the target language to provide them with more clues on how to use the expression. The (4) in Fig. 2 provides an example, where the NNSs select *"cuisine"*, ChatLearn provides an explanation in Chinese: *"Cuisine ... 菜系"*, followed by offering another example: *"I love Japanese cuisine"*.

Expression Extractor identifies L1-scaffolded input segments (i.e. Chinese text) as expression difficulties. These challenging parts are underscored and explained to attract NNSs' attention. This design is grounded in the noticing hypothesis [71] and output hypothesis [79], drawing NNSs' attention to the expression gaps that they find challenging to construct. For instance, if an NNS types *"There are many 美食 in Chongqing, especially 麻辣火锅"*, the corresponding part with Chinese characters will be recognized and underscored in the translation ("*There are many underlines in Chongqing, especially mala hotpot*") with an explanation (美食: *cuisines,* 麻辣火锅: *mala hotpot*). The implementation detail of this feature is shown in Appendix Table 6. Firstly, from the user's input, *extract* meaningful expressions that the NNS wrote in their Chinese, that would be valuable for them to learn in English. Secondly, *translate* the entire message into English language. Lastly, in the translated English message, *find and map* the parts that correspond to the Chinese expressions extracted earlier. This ensured that the appropriate parts of the translated English message were appropriately highlighted to the user and linked to each original Chinese expression.

**Table 1: Comparison of support components between *Baseline* and *ChatLearn* systems**

| System | Comprehension Support | Expression Support | Learning Support |
|---|---|---|---|
| **Baseline** | Comprehension Assistance: Translates received English messages into the NNS's first language. | Expression Builder: Allows users to type messages partially or entirely in their first language and translate into English. | None. |
| **ChatLearn** | Comprehension Assistance (same as Baseline). Expression Explorer: Enables users to manually select content for translations and examples. | Expression Builder (same as Baseline). Expression Extractor: Automatically detects scaffolded first-language segments in user input and explains them. | Expression Explorer: Transforms comprehension challenges into learning opportunities. Expression Extractor: Transforms expression challenges into learning opportunities. Contextual Review Cards: Recalls previously encountered expressions for spaced and incidental learning. |

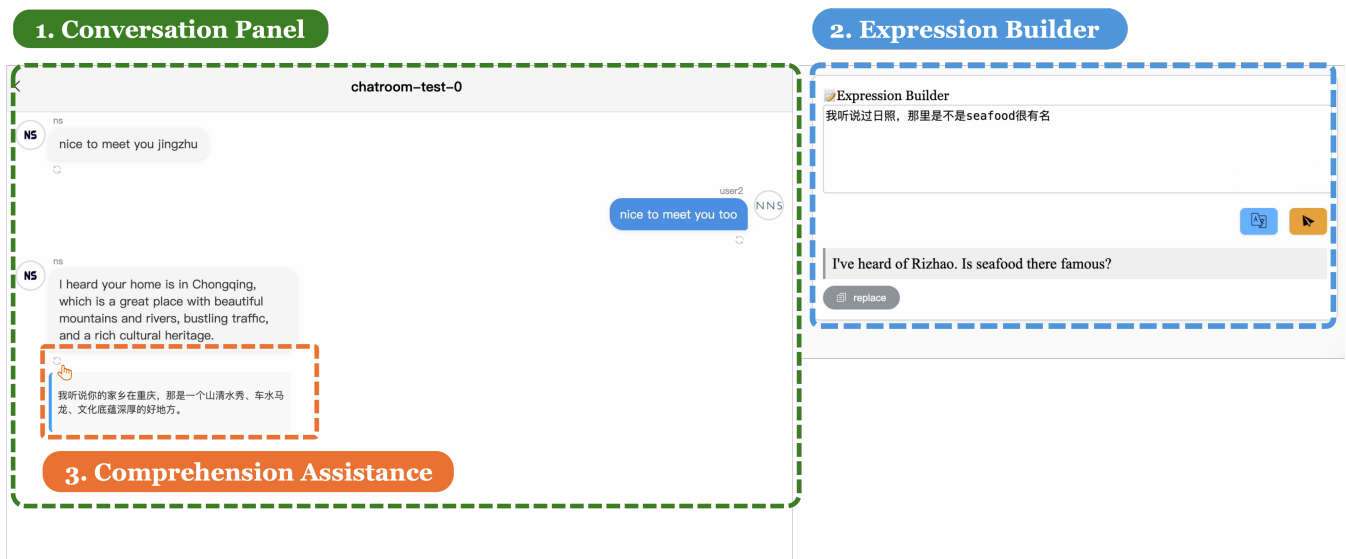

**Figure 1: Baseline interface. The system provides only communication-oriented support, including: (1) Conversation Panel for displaying chat history, (2) Expression Builder for constructing responses with NNS's native-language as scaffolds to generate translation, and (3) Comprehension Assistance for translating NS messages into the NNS's native language.**

***G2: Embed Incidental Learning in an Authentic Context.***
We also aimed to leverage the act of conversation itself as a valuable resource. In line with incidental learning [23] and spaced repetition [39], we wanted to highlight repeated instances of challenging expressions appearing over the course of conversation, to improve recall and retention. Further, inspired by theories of contextual learning [6] and language socialization [56], we aimed to emphasize the social and contextual interactions in which these challenging expressions previously appeared (Section 2.2.2). Therefore, we designed *(6) Contextual Review Cards*, which surface repeated examples of challenging expressions in a relevant context. When users translated expressions using the Expression Explorer or Extractor, as detailed above, we recognized these as potentially challenging phrases. We embedded these with OpenAI embedding models (text-embedding-3-large[4]) and stored them in a vector database (Chroma[5]). Subsequently, when users encountered a similar expression again, we retrieved and displayed a corresponding review card containing the previous use of the expression. These cards were triggered in two scenarios. *(6-a) Context-driven recall*: When a new message in English arrived, it was matched against all entries in the database; the top 3 most semantically similar saved expressions were then shown to users[6]. *(6-b) Expression-driven recall*: When NNSs were building expressions with AI assistance, stored expressions that were semantically similar were retrieved and displayed in the same manner.

---

[4]https://platform.openai.com/docs/models/text-embedding-3-large
[5]https://docs.trychroma.com/docs/collections/add-data

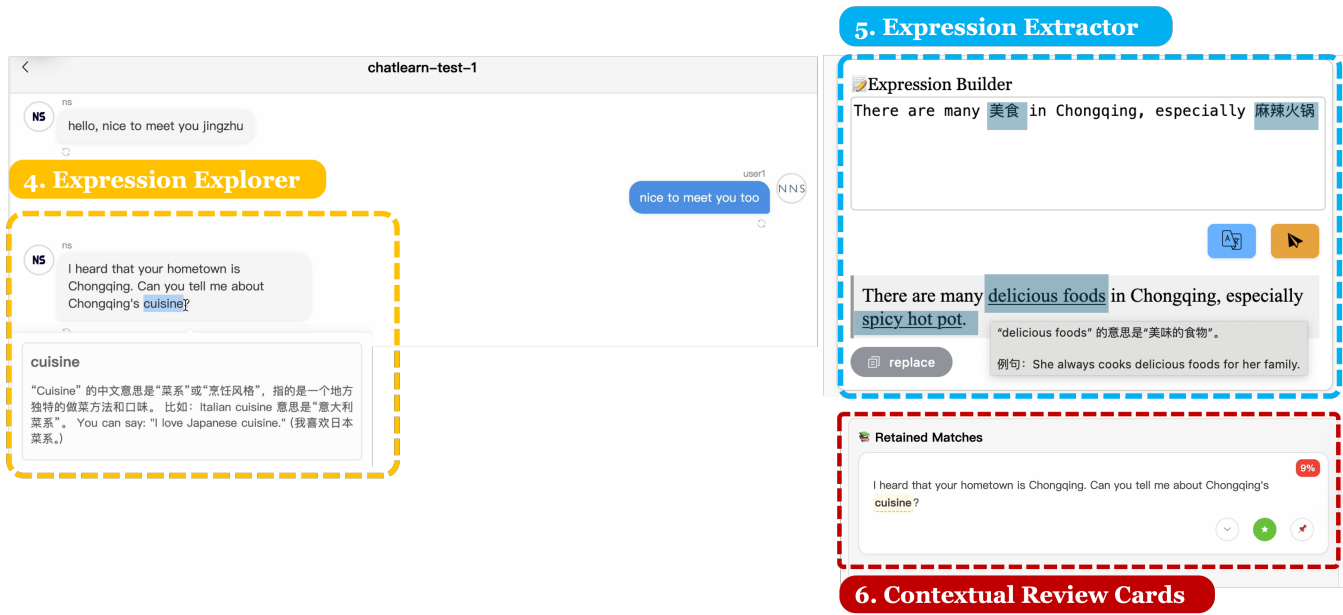

**Figure 2: ChatLearn interface. In addition to the baseline functions, ChatLearn integrates learning-oriented features: (4) Expression Explorer for actively noticing and querying expressions, (5) Expression Extractor for extracting and mapping scaffolded native-language input to target-language expressions, and (6) Contextual Review Cards for recalling and revising previously encountered expressions based on conversation context.**

This allowed review cards to appear intermittently over the course of conversation (either writing or receiving messages), allowing them to be displayed to the user in a time-spaced manner [10, 60]. Rather than only containing the isolated expression, each review card included the specific text/message surrounding the expression when it was saved, leveraging the specific context of its use to help the user better grasp and memorize the expression [26, 75]. For instance, the review card (see Fig. 2 (6)) displayed shows *"I heard that your hometown is Chongqing. Can you tell me about Chongqing's cuisine"* rather than only *"cuisine"*.

***G3: Balance Communication and Learning Goals.*** Finally, cognizant of how language learning might hinder the already challenging process of second-language communication (Section 2.3), we designed ChatLearn's features to create as little additional cognitive load as possible. Both the *Expression Explorer* and *Expression Extractor* enable learning to emerge implicitly from users' natural communication behaviors. Users are guided to actively notice translated expressions, but without having to consciously search for learning material. These features therefore unobtrusively transform comprehension and expression challenges into personalized learning opportunities. Meanwhile, the *Contextual Review Cards* further support acquisition by resurfacing previously encountered expressions in relevant conversational moments and at relevant timings of expression construction, providing incidental [23, 33], spaced [39], and contextualized review [26, 75, 82]. These cards appear automatically, without requiring the user to remember to bring them up; the contextual mechanism also ensures that only expressions relevant to the user's current actions (i.e., reading messages or writing expressions) are surfaced, avoiding irrelevant distractions that detract from the present task.

---
[6]To avoid displaying irrelevant expressions, we only displayed matches with a cosine similarity of >=0.15, a threshold determined through preliminary testing.

Taken together, these mechanisms are tightly integrated into the communicative flow, requiring minimal additional tasks and no switching between interfaces, thereby preventing extra obstacles to smooth and meaningful NNSNS interaction.

## 3.2 Participants

We recruited 44 pairs of participants (88 individuals) from the local community. Each pair consisted of one NS and one NNS whose first language was Mandarin Chinese. Pairs were randomly assigned to either the *Baseline* or *ChatLearn* condition. One pair in the *Baseline* condition was excluded post-experiment due to incomplete system logs, resulting in a final sample of **43 pairs** (86 individuals): 21 pairs in *Baseline* and 22 in *ChatLearn*.

Among these NNSs, 8 had a high school diploma, and 35 had a university or higher degree; 36 were male, and 7 were female; and their mean age was 23.26 (SD = 2.56). NNSs reported their English proficiency using: (a) self-rated English level on a 7-point scale (1 = Poor; 7 = Native-like), and (b) standardized test scores (including CET-4, IELTS, or TOEFL, when available). Their self-reported proficiency was $M = 3.09; SD = 0.68$. Standardized test scores of TOEFL[7], IELTS[8], and CET4[9] were converted into CEFR-equivalent levels (A1–C2) using concordance tables. CEFR levels were then coded on a 16 numeric scale (A1=1, A2=2, B1=3, B2=4, C1=5, C2=6). NNSs standardized score was: $M = 4.070; SD = 0.737$.

To verify that random assignment produced comparable NNS groups despite unequal final sizes, we conducted Welch's t-tests [90] and found no significant group differences in education level ($t(37.72) = 0.67$, $p = 0.509$), self-reported language ($t(40.05) = -0.46$, $p = 0.647$), and CEFR-standardized score ($t(40.86) = -1.02$, $p = 0.313$) between the two groups of NNS participants. All procedures were approved by the National University of Singapore Institutional Review Board (NUS IRB), and all participants provided informed consent prior to the study.

### 3.3 Procedure

#### 3.3.1 Conversation Topic and Flow.
Participants were instructed to discuss issues regarding social media use in modern society. We chose this topic for its prevalence in daily life [3], ensuring that as many participants as possible would be familiar with the topic and be able to contribute actively to the discussion. We instructed NS participants to lead the conversation by posing questions according to a discussion guide (see Appendix Table 7) that we provided them. This was to simulate realistic scenarios where native speakers often feel more confident in expressing their ideas and viewpoints, while preserving consistency in conversation flows across participants. We also instructed NSs to treat the conversation as a regular discussion about societal issues, rather than as an interview or test. The conversation took place entirely in English.

To preserve ecological validity and focus on self-directed behavior, we did not require NNSs to complete any explicit learning tasks during the conversation. Instead, learning features in ChatLearn were self-regulated and non-intrusive, allowing users to engage with them based on personal intention. This design allows us to examine whether learning behaviors can emerge naturally in communication and how existing functions may influence such behaviors.

#### 3.3.2 Overall Study Flow.
Before the study began, all NNS participants completed a *(1) pre-survey*, and were randomly assigned to one of two conditions (Baseline or ChatLearn) as well as a unique timeslot. During the study itself, we invited participants to a Zoom call, and assigned the NNS and NS to different breakout rooms for *(2) experiment preparation*. The NS was briefed on the discussion guide (as mentioned above), while the NNS viewed an instructional video on the system's functions (either *Baseline* or *ChatLearn*). Next, participants proceeded to the *(3) official experiment* with the NS and NNS chatting with each other. NNSs were instructed to disable browser extensions related to writing assistance, such as Grammarly and ChatGPT. This lasted for around 60 minutes.

After the conversation, the NNS completed a short *(4) free-recall test* (3 minutes long), where they recalled newly learned expressions during the conversation. Finally, all participants completed a *(5) post-survey*. The NNS answered questions measuring their self-rated learning motivation, tool usage experience, language acquisition performance, and communication experience during the task. For the NS, only the communication experience was relevant and thus measured.

Additionally, we conducted a *(6) post-study interview* with NNSs who expressed willingness and interest. This lasted around 20 minutes and was conducted directly after the main experiment.

### 3.4 Measurements

We used three types of measurements: system log data, post-surveys, and semi-structured interviews.

#### 3.4.1 Initial Learning Motivation (RQ1).
We adapted the *intrinsic* and *extrinsic* motivation subscales from the Language Learning Orientations Scale (LLOS) [55], which builds upon self-determination theory (SDT) [18]. The adapted items included two items to measure intrinsic motivation: *"I enjoyed discovering new words and expressions during the conversation"*, *"I found it interesting to learn phrases and words in different contexts"*. It also included two items for extrinsic motivation measurement: *"I hope to communicate more fluently with others in English"*, *"I believe improving my English is beneficial for my personal growth"*. All items used a 7-point Likert scale (1 = Strongly disagree, 7 = Strongly agree). The average score served as the measure of the **initial learning motivation**. The adapted scale demonstrated acceptable internal consistency (Cronbach's $\alpha = 0.759$). Similarly, we separately calculated the average scores of subitems in the intrinsic motivation and extrinsic subscales as the representative value of **initial intrinsic motivation** (Cronbach's $\alpha = 0.787$) and **initial extrinsic motivation** (Cronbach's $\alpha = 0.790$).

#### 3.4.2 Cognitive Load (RQ1).
We adapted the Cognitive Load Questionnaire (CLQ) [44], drawing from two of its subscales: **extraneous cognitive load**, to capture cognitive load caused by system-related distractions, and **germane cognitive load**, which conversely reflects productive mental effort invested in learning and meaning construction. Both were a 7-point Likert scale (1 = Strongly disagree, 7 = Strongly agree). Extraneous load items included: *"I had to put a lot of effort into figuring out how to use the system"*, *"The system's interface was distracting or confusing"*, *"Using the features of the system interrupted the flow of communication"*, and *"I was overwhelmed by the number of options or pop-ups presented"*. Germane load items included: *"I spent time thinking about the meaning of new expressions"*, *"I tried to explore and figure out more about these new expressions"*, *"I paid special attention to unfamiliar phrases that seemed useful"*, *"I tried to actively reflect on the expressions I wanted to remember"*, and *"When I recognized expressions I had seen before, I spent time to internalize them"*. For each measure, the average score of the items was calculated to represent the measure. Their internal consistency scores were: Cronbach's $\alpha = 0.804$ for extraneous cognitive load, and Cronbach's $\alpha = 0.852$ for germane cognitive load.

*3.4.3 Recall Performance (RQ1).* To assess participants' internalization of expressions encountered during communication, we conducted a free-recall task [69], commonly used to evaluate the quantity of content that the user can actively recall without prompting [8, 37, 88]. While prior studies implemented this test using unprompted meaning [88] or multiple-choice questions [37], our task contained no pre-defined set of target learning words to draw from. Therefore, we used a more open-ended method of asking participants to list as many expressions as they could within a 3-minute period.

To ensure the reliability of recall evaluation, we defined a *valid recalled expression* as any English word or phrase that (a) appeared during the conversation, either in a message from the NS counterpart or in system-generated translation; and (b) was newly noticed or learned by the NNS participant. Variants conveying the same meaning, such as *"privacy invasion"* and *"invasion of privacy"*, were counted as a single expression. All recalled items were manually checked.

When evaluating the recall performance, each participant needed to (1) recall expressions that they newly learned in the conversation. They then rated each expression based on (2) confidence, defined as how confident they would feel using such an expression in daily communication (7-point Likert scale; 1 = Not confident at all, 7 = Extremely confident), and (3) difficulty, representing how hard they considered the expression to use or remember (7-point Likert scale; 1 = Not difficult at all, 7 = Extremely difficult).

From these indicators, we calculated the measurements of *(1) recall quantity, (2) recall confidence*, and *(3) recall difficulty* of the *Baseline* and *ChatLearn* conditions, respectively. *Recall quantity* was defined as the total number of valid expressions each participant successfully recalled. *Recall confidence* was computed as the average confidence rating across all valid recalled expressions, and *recall difficulty* was calculated analogously as the average perceived difficulty of the recalled expressions.

*3.4.4 Perceptions of Language Acquisition (RQ1).* In addition to objective recall performance, we evaluated participants' perceptions of their own language acquisition performance. We developed a three-item scale, adapted from the general Task Performance Self-Assessment Framework (TPSAS) [70] to our specific tasks. The items used a 7-point Likert scale (1 = Strongly disagree, 7 = Strongly agree) and included: *"I believe that after this experiment, my language knowledge has increased"*, *"I think I performed well in remembering some knowledge, such as vocabularies, expressions, phrases during this experiment"*, *"I have indeed encountered many valuable knowledge, such as vocabularies, expressions, phrases throughout the entire task, but I feel that I struggled to remember and internalize them very well"* (reverse-coded). The average score of the items was calculated to represent the language acquisition performance. The scale demonstrated acceptable internal consistency (Cronbach's $\alpha = 0.743$).

*3.4.5 Logs of User Behaviors and System (RQ2, RQ3).* For both *Baseline* and *ChatLearn*, we recorded NNS' *(1) expression support count*, the number of times a participant asks for AI assistance to support their self-expression. We also calculated their *(2) first-language usage ratio*: when NNSs requested expression support, this was calculated as the number of first-language tokens input divided by the total number of tokens input.

With comprehension, we measured the number of times a participant requested translations of NS's messages. Both sets of participants had the ability to translate entire received messages, which we recorded as the *(3-a) full comprehension support count*. *ChatLearn* participants also had the option of manually selecting only parts of the message for translation. We recorded these as the *(3-b) partial comprehension support count*.

Additionally, we recorded the following unique behaviors unique to the *ChatLearn* condition. *(4) Learning opportunities* was the number of expressions that were obtained through manual expression selection and the automatic expression extractor. *(5) Review card interaction count* reflected the interaction frequency of NNSs with the retrieved review cards. *(6) Review card trigger frequency*, which was evaluated by how many times a card is automatically triggered for recall by the program throughout the task.

Besides, we also counted the *(7) message count* and *(8) message length*, calculated by the number of tokens, of NNSs in communication.

*3.4.6 Communication Experience (RQ2).* To evaluate the impact of AIMC tools on participants' communication experience, we adapted the Quality of Communication Experience (QCE) scale [48]. This scale included three dimensions. *(1) Clarity* assessed how clearly participants felt they could understand others (**comprehension clarity**) with item: *"I can easily understand what my partner was saying during the conversation"*, and express themselves in the group conversation (**expression clarity**). Items included: *"I believe my partner clearly understood what I was trying to express"*. *(2) Responsiveness* captured how smooth and timely the conversation felt, and whether the tool introduced any interruptions or cognitive burden. The items are: *"I felt that my responses in the conversation were timely and appropriate"*, *"I feel that using this tool can lead to noticeable pauses in the conversation"* (reversed-coded), *"The conversation with my partner was smooth"*. *(3) Comfort* measured participants' perceived psychological comfort and naturalness during the interaction. The items include: *"I felt nervous and pressured during the conversation with my partner"* (reversed-coded), *"I felt comfortable in the conversation with my partner"*. All items were rated on a 7-point Likert scale (1 = Strongly disagree, 7 = Strongly agree). The average score of the items was calculated to represent the communication experience. Their internal consistency is: Cronbach's $\alpha = 0.766$.

*3.4.7 Semi-Structured Interview.* We conducted follow-up semi-structured interviews to obtain complementary qualitative insights into participants' underlying cognitive and motivational processes. We invited all participants, and finally 30 NNS participants opted in (21 from *ChatLearn* and 9 from *Baseline*). The interviews were conducted in their native language, Mandarin Chinese, to allow the NNSs to express themselves more freely and accurately. Interviews lasted approximately 20 minutes and were audio-recorded and transcribed with consent, before being translated into English by researchers fluent in both languages. We mainly interviewed NNSs in the *ChatLearn* condition to understand how its unique features affected learning and user experience. We started by asking how ChatLearn influenced their learning processes, e.g., *"In what ways did this system affect your learning of English?"*. If they indicated that ChatLearn enhanced their learning processes, we asked, *"What kinds of content did ChatLearn help you learn?"*. We also further asked which features of ChatLearn were the most useful, including how and when they used these features. Finally, to understand potential negative impacts, we asked questions about the overall impact on the communication experience, such as *("How do you think this tool influences your communication?")*. We also selected a portion of *Baseline* NNSs according their recall performance in the free-recall test, to learn about the circumstances under which they exhibited learning behavior without ChatLearn's features, with questions such as, *"In what ways did this system affect your learning of English?"*.

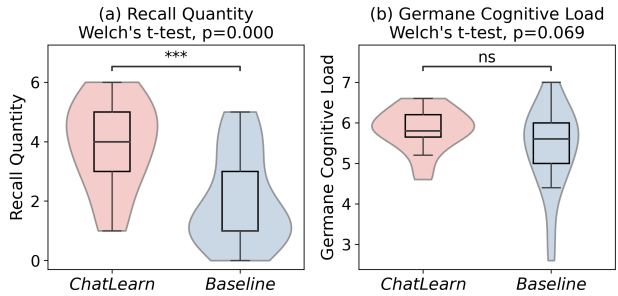

**Figure 3: Welch's t-test results on recall quantity and germane cognitive load. Violin plots show the data distribution, and box plots indicate median and interquartile ranges. Significance levels are marked as *** for p<0.001 and ns for non-significant. (a) Recall Quantity: Participants in *ChatLearn* recalled significantly more expressions than those in *Baseline*. (b) Germane Cognitive Load: Participants in the *ChatLearn* condition showed a trend toward higher germane load than those in the *Baseline* condition, although the difference was not significant.**

Two researchers conducted and independently coded the interviews using thematic analysis [9], resolving discrepancies in coding through discussion. This aimed to identify recurring themes related to motivation, learning behavior, system interaction patterns, and user preferences.

## 4 Results

In this section, we present (1) the outcome and mechanism of NNSs' language acquisition, (2) NNSs' communication experience and behaviors, and (3) NNSs' interactions with ChatLearn. We used JASP[10] to analyze the data. For between-condition comparisons on NNSs' behaviors and perceptions, we employed Welch's t-test, which accounts for potential unequal sample sizes or variances [90]. We subsequently conducted linear regression and path analysis to explore which factors and mediators can directly or indirectly affect recall performance. For convenience, we summarize the main results in Table 2.

### 4.1 Language Acquisition (RQ1)

Overall, *ChatLearn* participants recalled more expressions, while no statistically significant differences were found in germane cognitive load and extraneous cognitive load.

*4.1.1 Recall Performance.* In the free-recall test, the Welch's t-test showed that there was a significant difference in **recall quantity** of expressions ($t(40.72) = 3.97$, $p < 0.001$, $d = 1.211$), shown in Fig. 3(a). Participants in *ChatLearn* condition recalled $M = 3.909$; $SD = 1.509$ expressions, while those of participants in *Baseline* were $M = 2.048$; $SD = 1.564$. We found no effect of participants' English proficiency on their ability to retain language expressions. Specifically, we conducted two separate ANCOVA analyses using (1) self-reported English proficiency and (2) standardized CEFR-based proficiency (converted from CET-4, IELTS, and TOEFL scores) as covariates. Neither self-reported English proficiency ($F(1, 40) = 0.763$, $p = 0.388$) nor CEFR-standardized language level ($F(1, 40) = 0.047$, $p = 0.829$) significantly predicted recall performance.

The difficulty of recalled expressions was slightly higher for *ChatLearn* participants, though not statistically significant ($M = 3.54$, $SD = 0.72$ vs $M = 2.39$, $SD = 1.52$; $t(29.43) = 1.76$, $p = 0.088$, $d = 0.541$). There was no significant difference in perceived confidence in using recalled expressions ($t(32.76) = 1.16$, $p = 0.256$, $d = 0.355$).

*4.1.2 User Perceptions.* The Welch's t-test did not detect a significant difference in *germane cognitive load* between the two conditions ($t(29.268) = 1.885$, $p = 0.069$, $d = 0.579$), shown in Fig. 3(b). Despite this, we observed a trend that it was higher in participants who used *ChatLearn* ($M = 4.12$, $SD = 0.88$) than those in the *Baseline* condition ($M = 3.47$, $SD = 1.21$). Meanwhile, extraneous cognitive load did not differ significantly ($t(36.45) = 1.05$, $p = 0.301$, $d = 0.318$). There were also no significant differences in perceived language acquisition performance ($t(33.08) = 1.25$, $p = 0.219$, $d = 0.385$).

*4.1.3 Factors Influencing Learning.* To identify the factors that best explain **recall quantity**, we constructed a multiple regression model informed by prior theories in second language acquisition. The model included *learning motivation*, which has long been recognized as a fundamental determinant of second language learning success [62]; *extraneous cognitive load* that hinders learning and *germane cognitive load* that facilitates deeper schema construction and transfer, following cognitive load theory [44, 80]; and *tool usage factors* (e.g., review card interactions frequency, comprehension support count, and expression support count), as learning research suggests that the way learners interact with supports can shape both immediate performance and skill development [64, 66, 84, 96]. The experimental condition (*Baseline* vs *ChatLearn*) was also included as a factor.

The regression analysis showed that the full model ($M_1$), which included cognitive load variables (extraneous and germane), usage metrics, experimental condition, and initial motivation scores, significantly explained variance in recall quantity, $R^2 = 0.442$, $Adj$. $R^2 = 0.310$, $F(8, 34) = 3.361$, $p = 0.006$. This represents a substantial

**Table 2: Summary of significant quantitative findings.**

| Measure | Outcome | p-value | Effect Size |
|---|---|---|---|
| Recall Quantity | ChatLearn > Baseline | < 0.001*** | $d = 1.211$ |
| Recall Quantity (ANCOVA, self-reported proficiency as the covariate) | ChatLearn > Baseline | < 0.001*** | $\eta^2 = 0.283$ |
| Recall Quantity (ANCOVA, CEFR-standardized score as the covariate) | ChatLearn > Baseline | < 0.001*** | $\eta^2 = 0.276$ |
| Expression Clarity | Baseline > ChatLearn | 0.003** | $d = 0.975$ |
| Responsiveness | Baseline > ChatLearn | 0.039* | $d = 0.654$ |
| Message Frequency | Baseline > ChatLearn | 0.040* | $d = 0.650$ |
| Message Token Count | Baseline > ChatLearn | 0.034* | $d = 0.685$ |

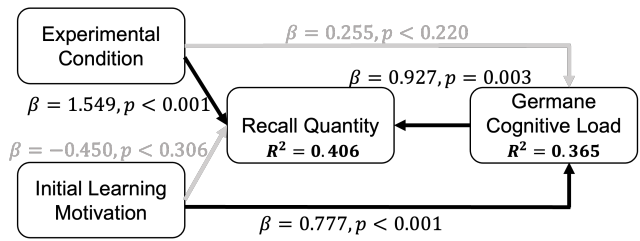

**Figure 4: Structural path model examining how initial learning motivation and experimental condition predict recall performance through germane cognitive load. Black lines and arrows represent significant paths: initial motivation positively predicted germane cognitive load ($\beta = 0.777, p < 0.001$), and germane load positively predicted recall quantity ($\beta = 0.927, p = 0.003$). The experimental condition (ChatLearn vs. Baseline) exerted a significant direct effect on recall ($\beta = 1.549, p < 0.001$). Gray lines and arrows represent non-significant paths. The model explained 36.5% of the variance in germane cognitive load and 40.6% of the variance in recall quantity.**

**Table 3: Multiple regression predicting recall quantity.** *$p < 0.05$, **$p < 0.01$, ***$p < 0.001$. **Germane cognitive load and experimental condition were significant positive predictors.**

| Predictor | $\beta$ | SE | t | p-value | VIF |
|---|---|---|---|---|---|
| (Intercept) | | 2.796 | -0.668 | 0.508 | |
| extraneous cognitive load | 0.079 | 0.338 | 0.508 | 0.615 | 1.457 |
| **germane cognitive load** | **0.444** | **0.348** | **2.721** | **0.010**** | **1.623** |
| total card interactions | -0.042 | 0.055 | -0.284 | 0.778 | 1.348 |
| comprehension support count | -0.088 | 0.011 | -0.529 | 0.600 | 1.689 |
| expression support count | 0.158 | 0.017 | 0.873 | 0.389 | 1.997 |
| **experimental condition** | **0.469** | **0.598** | **2.771** | **0.009**** | **1.744** |
| initial intrinsic motivation | 0.047 | 0.317 | 0.307 | 0.761 | 1.458 |
| initial extrinsic motivation | -0.213 | 0.509 | -1.201 | 0.238 | 1.908 |

Model fit: $R^2 = 0.442$, Adj. $R^2 = 0.310$, $F(8, 34) = 3.361$, $p = 0.006$

#### 4.1.4 Mediation Analysis.

To further understand what factors may contribute to variations in germane cognitive load, we conducted a Pearson correlation analysis. The results showed that **germane cognitive load** was significantly associated with both **initial intrinsic motivation** ($r = 0.457, p = 0.002$), **initial extrinsic motivation** ($r = 0.508, p < 0.001$), and **initial learning motivation** ($r = 0.586, p < 0.001$). These results suggest that learners who reported higher initial motivation–both intrinsic and extrinsic–tended to invest more germane cognitive load during the interaction.

Given the impact of initial motivation on germane cognitive load (initial learning motivation -> germane cognitive load), and germane cognitive load as an important factor influencing recall quantity (germane cognitive load -> recall quantity), to further clarify the mechanism, we conducted a mediation analysis, using *initial learning motivation* as the predictor, *germane cognitive load* as the mediator, and the outcome being *recall quantity*. To simultaneously consider the impact of *experimental condition* on germane cognitive load and recall quantity, we also included it as a predictor.

The structural path model (see Fig. 4) demonstrated satisfactory explanatory power. **Germane cognitive load** was significantly predicted by **initial learning motivation** ($\beta = 0.777, p < 0.001$), explaining 36.5% of its variance ($R^2 = 0.365$). In turn, **germane cognitive load** exerted a strong positive effect on **recall quantity** ($\beta = 0.927, p = 0.003$).

The **experimental condition** (ChatLearn vs. Baseline) had a significant direct effect on **recall quantity** ($\beta = 1.549, p < 0.001$), improvement over the intercept-only model ($M_0$), which explained no variance.

Among the predictors, both **germane cognitive load** ($\beta = 0.444$, $t = 2.721$, $p = 0.010$) and **experimental condition** (using *ChatLearn* or *Baseline*) ($\beta = 0.469$, $t = 2.771$, $p = 0.009$) emerged as significant positive predictors, indicating that participants who experienced higher germane load and those assigned to use *ChatLearn* recalled more expressions. Other predictors–including extraneous cognitive load ($p = 0.615$), intrinsic/extrinsic motivation ($p = 0.761$, $p = 0.238$), and usage metrics–were non-significant.

but its indirect effect via germane cognitive load was non-significant ($p = 0.257$). Likewise, the direct effect of initial motivation on recall was non-significant ($\beta = -0.450$, $p = 0.306$), although its indirect effect through germane load reached significance ($\beta = 0.720$, $p = 0.014$). Overall, the model explained 40.6% of the variance in recall quantity ($R^2 = 0.406$).

These findings suggest that learners' initial motivation enhances recall performance primarily *indirectly*, by increasing their germane cognitive engagement during the communication task. In contrast, ChatLearn improves recall *directly*. Thus, learner-driven cognitive engagement and system-driven design features contribute to expression retention—but through distinct psychological pathways.

## 4.2 Communication Experience and Behaviors (RQ2)

The results indicate that NNSs in *Baseline* had a better communication experience, especially in terms of **expression clarity** and **responsiveness**. However, NS did not believe that the communication experience caused by NNS in the two conditions had a significant difference.

### 4.2.1 Communication Experience.
The Welch's t-test showed that there was no significant difference between the two conditions ($t(36.89) = -1.84$, $p = 0.075$, $d = -0.562$). When examining the specific sub-constructs of communication experience (see Fig. 5a, we found that *Baseline* participants reported higher **expression clarity** ($M = 6.14$, $SD = 0.73$ vs $M = 5.27$, $SD = 1.03$; $t(37.79) = 3.21$, $p = 0.003$, $d = 0.975$) and **responsiveness** ($M = 5.67$, $SD = 0.78$ vs $M = 5.20$, $SD = 0.66$; $t(39.24) = 2.14$, $p = 0.039$, $d = 0.654$) than *ChatLearn* participants. No significant differences were observed for comprehension clarity ($t(34.92) = 1.24$, $p = 0.223$, $d = 0.380$) or comfort ($t(39.50) = 0.11$, $p = 0.910$, $d = 0.035$).

From the NSs' perspectives, there was no significant difference in communication experience between participants who interacted with either *ChatLearn* or *Baseline* NNSs ($t(40.64) = 0.50$, $p = 0.623$, $d = 0.149$).

### 4.2.2 Message Frequency and Message Token Count.
The Welch's t-test showed a significant difference in **message frequency** between the two conditions ($t(38.83) = 2.13$, $p = 0.040$, $d = 0.650$). NNSs in the *Baseline* condition produced significantly more messages ($M = 48.71$, $SD = 22.48$) compared to those in the *ChatLearn* condition ($M = 35.32$, $SD = 18.55$). In terms of overall **message token count**, *Baseline* NNSs generated significantly more ($t(28.65) = 2.23$, $p = 0.034$, $d = 0.685$) tokens ($M = 663.67$, $SD = 354.52$) than *ChatLearn* NNSs ($M = 472.68$, $SD = 172.31$). These results were shown in Fig. 5b.

### 4.2.3 Communication Support-Seeking.
The Welch's t-test showed that there was no significant difference between two conditions ($t(30.53) = 0.89$, $p = 0.379$, $d = 0.274$) on total **comprehension support count**. Within *ChatLearn* specifically, we compared the use of full support (obtaining comprehension support for the full message) and partial support (obtaining support for a manually-selected expression). Notably, a paired-samples t-test (see Fig. 6a) revealed that the number of **full comprehension support** instances ($M = 20.93$,

$SD = 27.10$) was significantly higher than that of **partial comprehension support** ($M = 5.19$, $SD = 12.14$), $t(42) = 3.92$, $p < 0.001$, $d = 0.60$.

We found no significant differences between conditions for expression support count ($t(32.51) = 1.51$, $p = 0.140$, $d = 0.464$) and first language usage ratio ($t(35.72) = 1.61$, $p = 0.116$, $d = 0.493$).

## 4.3 Interactions with ChatLearn Features and Learning Opportunities (RQ3)

The descriptive statistics showed that when using ChatLearn, participants produced $M = 33.91$ different expressions ($SD = 35.98$). In the context of NNS-NS dialogue, each expression was triggered an average of 4.95 times ($M = 4.95$; $SD = 6.75$). The following result indicates that more learning opportunities are produced when NNSs seek to construct expressions. The paired-samples t-test showed that the **learning opportunities** produced through comprehension support and expression support were significantly different ($t(42) = -3.41$, $p = 0.001$, $d = -0.52$), as shown in Fig. 6b. The instances of expression support ($M = 12.16$, $SD = 20.33$) were significantly more numerous than those produced through comprehension support ($M = 5.19$, $SD = 12.14$). Additionally, participants engaged in an average of $M = 4.64$ interactions ($SD = 5.99$) with the review card.

To further explore which factors affect NNSs' adoption of comprehension support, we conducted a Pearson's correlation analysis. The results showed that the self-reported language proficiency is negatively correlated with both the number of comprehension support ($r = -0.306$, $p = 0.046$) and the number of expression support ($r = -0.416$, $p = 0.006$). This suggests that people who report less proficiency in their English tend to use the tools more frequently to support the communication process.

## 4.4 Qualitative Analysis (RQ2, RQ3)

Through qualitative analysis, we attempt to supplement quantitative data. On one hand, by revealing the strategies NNSs use with ChatLearn during communication, we explain why ChatLearn has no significant impact on communication and find that ChatLearn changes the attitude of using AI support (RQ2). On the other hand, we summarize the NNSs' perceptions of how ChatLearn captures learning content through noticing design and subtly promotes their acquisition through the repetition mechanism (RQ3).

### 4.4.1 Balancing Learning Motivation with Communication Priorities.
NNSs using ChatLearn tended to focus mainly on communication, while viewing learning as a secondary consideration. We found, however, that the presence of learning features created positive changes in NNS' behavior and attitudes in second-language communication.

**(1) An Acceptable Level of Distraction.** *ChatLearn* users reported experiencing mild distraction during conversations as they attended to learning new words and expressions, which aligns with the **lower responsiveness** observed in the quantitative results. As P38 explained, *"I spent some time learning; I was memorizing this phrase,"* and P46 noted that he was *"drawn in"* by the repeated word cards. Nevertheless, most users regarded this level of distraction as acceptable, describing it as *"not a big problem"* (P38) and something that *"didn't disrupt the flow of communication"* (P08). As P17 added,

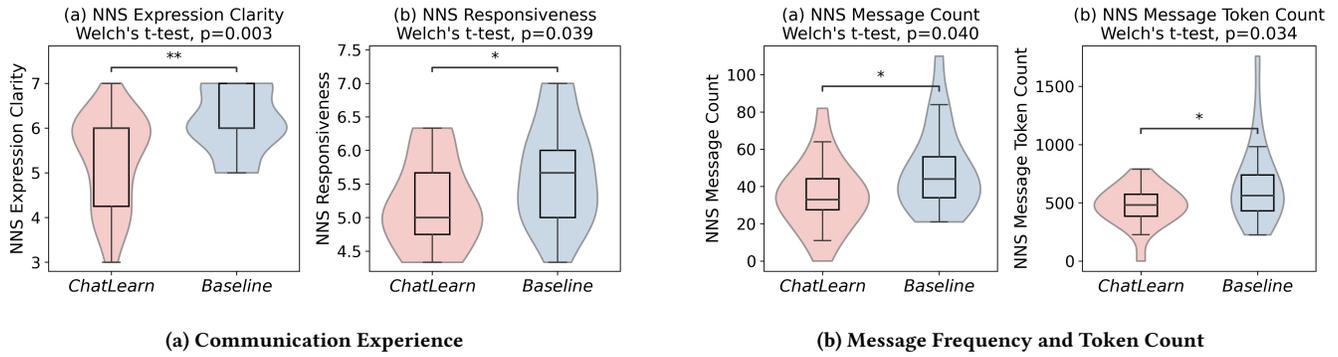

(a) Communication Experience      (b) Message Frequency and Token Count

**Figure 5: Welch's t-test results on communication experience and messages. Violin plots show the data distribution, and box plots indicate median and interquartile ranges. Significance levels are marked as \*$p<0.05$, \*\*$p<0.01$. (a) Communication Experience: _Baseline_ participants reported higher expression clarity and responsiveness than _ChatLearn_. (b) Message Frequency and Message Token Count: _Baseline_ participants produced significantly more messages and total tokens than _ChatLearn_.**

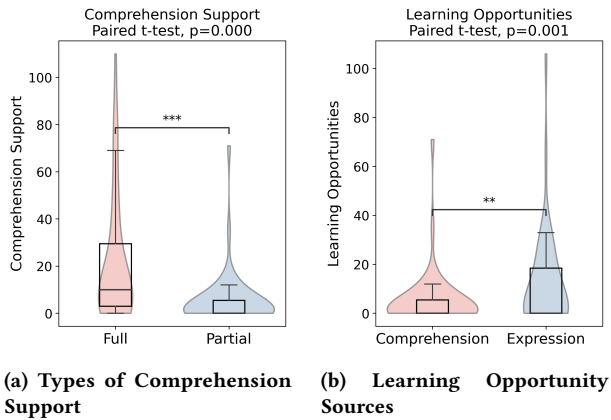

(a) Types of Comprehension Support    (b) Learning Opportunity Sources

**Figure 6: Paired-samples t-test results on communication comprehension support types (from full comprehension support or from partial comprehension support) and different source of learning opportunities (from comprehensions or expressions) in _ChatLearn_ condition. Violin plots show the data distribution, and box plots indicate median and interquartile ranges. Significance levels are marked as \*$p<0.05$, \*\*$p<0.01$, \*\*\*$p<0.001$. (a) Types of Comprehension Support: Participants used significantly more partial comprehension support than full comprehension support. (b) Learning Opportunity Sources: The number of learning opportunities from expression support was significantly higher than those from comprehension support.**

_"I also usually look up unfamiliar words [from the dictionary] during conversations,"_ further suggesting that the learning-oriented features did not meaningfully hinder their communicative experience.

_(2) Communication Still as the Top Priority._ Despite the integration of learning-oriented features, _ChatLearn_ users consistently prioritized communication. Several participants emphasized that when expressing themselves, they were _"less focused on learning and more on expressing their own opinions"_ (P06). P11 remarked, _"It_ was for the purpose of communicating with him at the time, not for learning content," and P12 stated, "I didn't intend to write particularly good sentences; as long as the other person could understand, that was fine." Additionally, some users indicated that they chose to ignore certain learning-oriented features during fast-paced interactions. Both P19 and P22 stated that they focused on communication and _"didn't specifically notice"_ the repeated word cards. P02 further elaborated, _"In fast-paced real-time chats, there might not be such opportunities [to seize learning opportunities]."_

_(3) Learning During Conversational Pauses._ To ensure the communication was not disrupted, NNSs chose to review the learned content during conversation gaps. P38 mentioned, _"During the conversation pauses, I looked at my pinned [word cards],"_ while P50 noted, _"While I'm waiting for a reply, I'll check my pinned cards."_ P33 further pointed out, _"I felt like the other person wasn't saying anything when I marked [word cards]."_

Some _ChatLearn_ users emphasize that communication might be a good way to accumulate content, but they postpone the learning process until the conversation ends. For example, P19 mentioned _"Learning during a conversation isn't very efficient. It's probably best to review the card after the conversation ends.",_ while P56 noted _"Because I may be familiar with it [during communication], but sometimes when I have free time, I want to check it again."_ Similarly, P08 also hoped the system to _"create a small notebook or bookmark... [for me]to review what I learned and asked in that conversation."_

_(4) Different Consideration of Using Expression Support._ Although the NNSs in ChatLearn did not view learning as a burden, some of them seemed to exhibit a different attitude towards AI content compared to the _Baseline_ group. Specifically, _Baseline_ users tended to accept the provided translations as-is. P57 only _"glanced through"_ the translations, while P54 noted they did not examine them in much detail. P02 commented, _"I remember I only looked at one or two, and I felt that the translations were quite good. I think I didn't pay much attention afterwards."_ In contrast, _ChatLearn_ users exhibited more caution when using this feature. P17 and P31 both noted that they tended to check the translated text before sending it, with P17 in particular saying that they _"used [ChatLearn] to check if [their] words were misspelled"._ P06 expressed similar sentiments, worrying that their inputs might have contained errors. This may explain the lower **expression clarity** reported by ChatLearn users in the quantitative results.

*4.4.2 How Interacting with ChatLearn Transforms Communication into Meaningful Learning Opportunities.* We identified three main themes for how NNS participants learnt new expressions through ChatLearn.

**(1) Noticing Unfamiliar, Interesting, and Native Expressions.** When users encountered novel or native-like expressions during dialogue, which they felt could not be learned from textbooks, NNSs will naturally pay attention to the meanings of these expressions. As one participant described: *"The other party is a native speaker. He uses some more local expressions. These vocabulary words might not be used during the learning process. After translation, I found it quite novel."(P26)* Some intriguing expressions will also attract NNSs to memorize automatically: *"When he brought the topic to 'brain rot'. I really have a deep impression of this word." (P11)*

In addition to focusing on the native and interesting part of the NS language, NNSs in *ChatLearn* condition also cared about whether the information they convey performed the same. P38 praised the system for *"expressing my ideas in a more native way"*, while P12 and P46 hoped the system could explicitly check *"whether an expression was correct"* from NS's opinion. P19 clearly noted *"If [the system] could tell me if this is how the locals say it, I think that would be quite helpful."* P28 elaborated on this concern, explaining, *"Because I was unsure whether 'Can you' or 'Do you' was better ... then [the system] told me 'do you' was better."* These concerns about their own content can be seen as a type of noticing initiated by NNSs, independent of ChatLearn's noticing design.

**(2) Expression Extractor and Explorer Distinguish Learnable Materials.** *ChatLearn* users found the **Expression Extractor** useful for learning. This was particularly true for mixed-language inputs where native-language (Chinese) expressions were mapped to their corresponding target language (English) translations. Some mentioned how conventional systems that translate entire messages without further highlighting made it easy to forget (P06) or overlook (P35) the expressions that they struggled with initially. P08 in particular noted how with other systems, *"I might have just finished the translation and moved on... there wouldn't be a subsequent process of learning"*. In contrast, *ChatLearn* was able to extract important phrases for learning (P22) and, by helping participants actively notice them (P08), encouraged forming deeper impressions of these expressions (P06, P35). A few participants also noted how the **Expression Explorer**, which allowed highlighting of specific words within received English messages for translation, aided learning. P11 mentioned how, if the entire message was translated from English to Chinese, *"I might still be searching within the Chinese text for what the word is. But if I highlight the words [in English that I am unfamiliar with], I can directly know what the word [in Chinese] is."*

The above results reflect the importance of noticing and the significance of features that moderately transform the NNSs' noticed content into learning opportunities. Some NNSs even suggest that, in addition to recognizing the difficult expressions, the system can also provide polishing to their content and highlight the differences between these polished parts and their original usage (P31, P22) to strengthen noticing so that they can simultaneously learn how to express themselves in a more native way.

**(3) Contextual Repetition through Review Cards.** The preemptive nature of how *ChatLearn* prompted users with review cards also aided spaced and incidental learning. P08 and P13 mentioned how the occasional appearance of the cards reminded them to review the meaning and context of certain expressions. The cards also provided users with a way of incidental learning. While they did not intentionally attempt to memorize the content of the cards (P13, P22), these cards still gave users opportunities to *"scan through"* (P22) or an applied use (P08, P13) expressions they had seen before, created *"repeated exposure"* (P22) and reminders (P08) that helped with learning without detracting from the main conversation task.

Additionally, participants noted a few other benefits of *ChatLearn* and suggestions for real-world implementations. P22 appreciated how the dialogue setting of our study provided **realistic usage examples of unfamiliar expressions**, unlike traditional methods in which they struggled to understand where and how to use such expressions. P08 added that *ChatLearn*'s **integration of language learning** into an applied use case (i.e. conversation) was particularly helpful, as opposed to having learning material in a *"separate app"* which created systemic friction in applying what they were learning. Meanwhile, P12 mentioned the importance of a **persistent list of saved expressions across multiple conversations** for real-world applications of ChatLearn, which would also allow for further revision outside of conversations.

## 5 Discussion

Our study shows the potential for ChatLearn to aid second-language learning through conversation, evidenced by the significant improvement in NNSs' recall of expressions (RQ1.1). Further analysis demonstrates how learning motivation and system usage are two main drives for these results, with germane cognitive load as the critical mediator (RQ1.2). At the same time, although overall communication experience was not significantly affected, using ChatLearn reduced perceived clarity and responsiveness of expression (RQ2). Finally, we revealed how learning opportunities mainly arose, especially when NNSs encountered novel NS expressions, struggled with their own constructions, or engaged with review cards (RQ3). Together, these findings highlight both the benefits and trade-offs of embedding language learning into AIMC systems. We discuss these findings and outline implications for the design of future AIMC systems to balance both communication and learning goals.

### 5.1 Design and Motivation: Factors Enhancing Learning Performance

Our regression and mediation analyses (Section 4.1.3 and 4.1.4) show that recall performance is influenced by two key mechanisms. On one hand, higher initial motivation promotes higher germane cognitive load, which *indirectly* helps users achieve better recall performance. This is consistent with previous research, which emphasizes the effectiveness of germane cognitive load in promoting learning behavior [17, 29]. Our results further emphasize the promotion of motivation as an important starting point on this path.

This suggests that AIMC systems should focus on improving learning motivation, thereby helping NNS learners actively shift from relying on AI to discovering learning opportunities with AI assistance.

On the other hand, the design of ChatLearn **directly** empowers NNSs, enabling them to recall more expressions through the combination of features such as noticing, contextual review, and spaced repetition. This is reflected in NNSs qualitative feedback that active noticing and repetition were key elements of their learning process (see Section 4.4.2). This suggests the effectiveness of ChatLearn's design and its initial design goals, i.e., incidentally promoting continuous language development during communication. Furthermore and crucially, this was not at the expense of increased system usage complexity through higher extraneous cognitive load. This aligns with a key design goal in learning technologies [17], addressing our third design goal of minimizing extraneous cognitive distractions in the learning process. Notably, while there was no significant difference in germane cognitive load (Section 4.1.2) between ChatLearn and the baseline, recall performance was still significantly stronger in the ChatLearn group. Therefore, future research can also move beyond objective-load and motivation perspectives, and investigate more advanced learning mechanisms relevant to improving the efficacy of AIMC and learning systems.

Finally, while objective recall performance improved, participants' subjective self-perceptions of acquisition performance did not differ significantly between the ChatLearn and Baseline conditions. Several NNSs also noted that their attention was more focused on communication than learning (Section 4.4.1). This suggests that NNSs in ChatLearn are likely to experience *incidental learning* [23, 33, 65], where learners acquire knowledge implicitly without necessarily perceiving the process or outcome. ChatLearn thus exemplifies how language acquisition can occur in the background of communication activities, demonstrating the effectiveness of such AIMC designs.

## 5.2 Communication vs. Learning: How Different Goals Interact and Co-Exist

ChatLearn did not significantly affect the overall communication experience (Section 4.2.1) while providing additional learning gains (Section 4.1.1). Despite this, participants in the *ChatLearn* condition did report lower levels of expression clarity and responsiveness. Our qualitative data provides an explanation for both these results, and suggest that they may not necessarily reflect negative effects on communication.

Regarding **expression clarity**, ChatLearn participants reported being more cautious and reflective when using expression support (Section 4.4.1), often checking translations for accuracy and appropriateness before sending. Since *ChatLearn* NNSs spend more time and effort crafting each message (Section 4.2.2), this may have naturally encouraged critical thinking about the content being sent. This has been found to reduce overconfidence in one's work, such as their written messages [91]; this may explain the reduced confidence in expression clarity that ChatLearn participants showed. In contrast, *Baseline* users tended to accept AI-provided translations with minimal scrutiny, *"glanc[ing] through"* them and expressing relatively higher confidence in the output clarity. This pattern aligns with automation bias [25], where users over-trust and under-verify automated decisions. It also resonates with concerns in the AIMC literature that the design of systems focused solely on efficiency can foster overreliance [32, 57], reducing users' critical engagement with the language and potentially stagnating long-term skill development [51], as well as decreased autonomy [92].

Therefore, the lower expression clarity in the ChatLearn condition may not reflect a design flaw, but rather a positive shift in their attitude to AI support, from passive recipients to critical collaborators with AI. Although this shift may temporarily reduce their confidence in output quality, in the long run, this critical engagement may be crucial for avoiding automation bias and promoting genuine language ability development. Future research can conduct long-term experiments to empirically investigate the effects of critical engagement with learning materials.

Regarding **responsiveness**, *ChatLearn* participants still indicated a relatively good responsiveness (M = 5.2 out of 7), despite this being lower than the *Baseline* condition. This disparity is not unexpected [64], as the ChatLearn system inherently contained interactive learning features not present in the Baseline system, which lead to multitasking [58]. However, the qualitative findings indicated that ChatLearn users tended to use these learning features to fill gaps in conversation in their own time (Section 4.4.1). These users did not specifically dedicate time and resources to fulfilling learning objectives, but instead engaged with learning features during time that would otherwise have been wasted. Even so, we acknowledge that this decrease in responsiveness may significantly impact communication in situations where fast-paced interaction is critical, such as online meetings [64]. As some NNSs described in interviews (Section 4.4.1), a feasible alternative is to accumulate learning materials during the conversation process and build subsequent acquisition mechanisms after the task is completed. Therefore, future research can investigate appropriate learning mechanisms tailored to specific communication scenarios.

Taken together, our findings with ChatLearn suggest an additional nuance to designs centered on communication efficiency. Specifically, when learning support is provided, NNSs may not focus solely on efficient communication; they may also actively engage in additional exploration and learning behaviors, attempting to achieve a balance among multiple objectives. Therefore, future AIMC research should investigate these other meaningful goals, such as skill development for users beyond communication goals.

## 5.3 Promoting Learning Through Noticing, Authentic Conversation, and Incidental Repetition

Overall, NNSs recognize that ChatLearn provides them with ample learning opportunities. Furthermore, individuals who self-assess as having lower language proficiency tend to seek more help (Section 4.3), which also suggests greater room for improvement and more learning opportunities from ChatLearn.

Regarding the features for learning, accumulating unfamiliar expressions through manual selection and automatic extraction during expression building helps them expose and highlight language deficiencies. Specifically, our participants noted the importance of active noticing [4, 71], which allows them to quickly distinguish unfamiliar parts from background noise. In contrast, traditional full translation methods make it difficult for them to identify unfamiliar expressions and commit them to memory (Section 4.4.2). Notably, NNSs gain the most opportunities through the expression phase. This is consistent with the output hypothesis [78, 79], emphasizing that constructing content helps discover one's deficiencies fully. This process encourages motivation for learning and to correct and compensate for these gaps. Therefore, future research on NNSs' language acquisition could focus more on the output construction and design more detailed support in this process.

Interestingly, NNS participants mentioned that the authenticity of their conversations with NS counterparts attracted them greatly. This differs significantly from conventional textbook learning, and aligns with claims [11] that NSs are viewed as *"the authoritative source of exemplary practice in the target language"*, which underlines why learners crave exposure to authentic language [34]. Such authenticity tends to motivate learners more than textbook-based practice. Future research could use LLMs to simulate authentic conversations at scale and promote NNSs' engagement and language development in such an environment [36].

Finally, NNSs place great importance on the repetition caused by review cards during the conversation. They emphasized that their learning occurred "incidentally" during conversation rather than through deliberate memorization. This aligns with research on *incidental vocabulary learning* [23, 65], which demonstrates that repeated exposure to meaningful input can lead to vocabulary gains even when learning is not the explicit goal. ChatLearn's contextual retrieval and spaced re-presentation of expressions operationalizes this mechanism, reinforcing exposure across conversational turns in a way that mirrors naturalistic language acquisition. This simple and effective mechanism could be emphasized in future research and widely applied in more communication scenarios to promote acquired behavior.

## 5.4 Design Implications

The ChatLearn system validates the feasibility of embedding language learning within real-time communication. Building on its core mechanisms and guided by the NNSs' behaviors and preferences observed in our study, we identify several key opportunities to advance the next generation of AIMC systems. The following design implications focus on enhancing learner motivation, elevating authenticity, and providing more intelligent, output-focused support to transform everyday communication into a more powerful and personalized language learning ecosystem.

*5.4.1 Facilitate Learners' Motivation through Personalized and Goal-Oriented Settings.* In our study, motivation was shown to significantly enhance germane cognitive load and finally improve recall performance. Motivation research [50] indicates that learning motivation largely depends on learners' perception of the task's value and its relevance to themselves. Therefore, future AI-assisted language learning systems should not only provide general support but also allow users to actively specify the specific expressions or scenarios they wish to improve. The system can then provide targeted support for these highly personalized goals, such as suggesting formal language or recommending authentic expressions. This "goal alignment" design essentially helps users build a sense of value and relevance for the task, fundamentally stimulating their motivation to learn.

*5.4.2 Support Noticing through Native-Like Scaffolding.* NNSs value authenticity and are drawn to native-like expressions, which was neglected when designing current translation-based supports [47, 57]. They often produce functionally correct but linguistically flat outputs. Future AIMC systems should provide scaffolding that mirrors native speaker phrasing, offering alternatives that are not only accurate but also stylistically appropriate. For example, instead of translating *"I think ..."* directly, the system could suggest *"From my perspective ..."* or *"It seems to me ..."* with brief explanations of usage context. This approach aligns with learners' desire for authentic input [11], making the communication support and learning experience more engaging.

*5.4.3 Enhance Output-Based Learning During Expression Construction.* The output phase proved to be a rich source of learning opportunities. Therefore, future research can build on the current design of ChatLearn in supporting NNSs' expression construction. For instance, systems can detect moments of hesitation or drafting pauses (e.g., through typing delay or backspacing) and offer immediate, context-aware suggestions–similar to how systems like Grammarly[11] surface real-time feedback. Instead of waiting for users to complete a sentence, the system could proactively highlight potential improvements or native alternatives, helping learners notice and correct gaps as they type. This real-time, output-focused approach strongly aligns with the output hypothesis [78, 79], turning every drafting moment into a learning opportunity.

*5.4.4 Integrate ChatLearn with AI-Simulated Authentic Conversations.* Authentic NS-NNS dialogue is a powerful motivator, but not always accessible. Currently, many AI chatbots simulate native speakers and are used in conversation practice to promote the language development of NNSs [2, 35, 36, 95]. Therefore, future research can combine these chatbots to simulate native speakers while using ChatLearn's mechanisms to further help NNSs convert language learning opportunities.

*5.4.5 Build Cross-Platform Learning to Reinforce ChatLearn.* Language development requires reinforcement across contexts and over time. Future AIMC systems can go beyond ChatLearn and function as a unified learning ecosystem that spans multiple platforms–chat, email, document editing, etc.–through browser extensions or integrated apps. Such an ecosystem could accumulate learning materials from all communication activities, creating a personalized expression library. Subsequently, when users write emails with AI [46, 49] or compose messages [12], the system could resurface relevant phrases or suggest previously learned vocabulary, encouraging continuous, contextual, and cumulative learning. This method also shifts the pressure of reviewing content to other scenarios that do not require an immediate response (e.g., saving expressions from real-time conversations and resurfacing them during asynchronous activities such as email writing).

---

[11]https://www.grammarly.com/

## 6 Limitation and Future Work

Our study has several limitations that warrant consideration. First, the experiment was conducted in a controlled setting with text-based communication between NNSNS pairs. Group or spoken communication may create more complex conversational dynamics and usability considerations. Future research investigates such richer communication settings to further validate and refine the design.

Second, the study involved Chinese NNSs communicating in English; while this provides valuable insights, whether and to what extent other languages can be acquired in a similar manner need further verification, due to the inherent differences between languages.

Third, our evaluation primarily relied on short-term recall and self-reported measures, which, although they demonstrated a potential direction, cannot fully reveal the long-term impact of ChatLearn on language development. Longitudinal studies are needed to determine whether the observed gains in recall translate into durable improvements in fluency and communicative competence.

## 7 Conclusion

This paper presented ChatLearn, an AIMC system that transforms NNSs' communication difficulties into learning opportunities. Across a mixed-methods study with 43 NNSNS pairs, ChatLearn improved expression recall without imposing substantial cognitive burden, though it reshaped users' perceptions of clarity and responsiveness. Our findings highlight how noticing, contextual resurfacing, and authentic conversational use can jointly support incidental language development during real communication. At the same time, the observed trade-offs underscore the challenge of integrating learning features into already demanding multilingual interactions. We propose design directions for AIMC systems that more deliberately balance communication fluency with opportunities for gradual skill growth. More broadly, this work illustrates how AI assistance can move beyond one-time support to foster ongoing language development within everyday conversation.

## Acknowledgments

This research was supported by the Ministry of Education, Singapore, under Grant A-8002610.

## A  Prompt Templates

Prompts for the various system features are listed below in Tables 4, 5, and 6.

## B  Discussion Guide for Native Speaker

Table 7 describes the instructions given to native speakers in both conditions.

**Table 4: Prompt template for translation, used for both comprehension assistance and expression assistance. Placeholders such as [USER_INPUT], [NATIVE_LANGUAGE], [TARGET_LANGUAGE], [CONTEXT] are dynamically replaced during runtime.**

---

**Instruction:**
Translate the user's message (in [NATIVE_LANGUAGE]) into fluent [TARGET_LANGUAGE].

**Requirement:**
- Consider the given context only to understand the situation.
- Do not include or translate the context itself in the output.
- Ensure the translation is natural and fluent.

**Input:**
Context: [CONTEXT]
User Message: [USER_INPUT]

**Output format:**
{"translated_text": "..."}

---

**Table 5: Prompt template for generating explanations, used by expression explorer. The placeholders within brackets [ ] are dynamically replaced with real-time data during the experiment. Specifically, [TARGET_LANGUAGE] is the language of explanation, [NATIVE_LANGUAGE] is the user's language, [PHRASE] is the selected expression, and [CONTEXT] provides conversation context information for disambiguation and better translation.**

---

**Background:**
You're an [TARGET_LANGUAGE] explainer for a non-native speaker.
The user is a non-native speaker.

**Requirement:**
Given the phrase: "[PHRASE]", explain it in [TARGET_LANGUAGE].
Then, try to slightly help the user internalize it. For example, you can give another daily example (in [NATIVE_LANGUAGE]) to demonstrate how to use the important expression in the phrase, or you can use other flexible ways to help.

- Be concise, because the user is in a real-time communication context.
- Context (if any): [CONTEXT or 'N/A'].
- Note: The context is only for understanding the situation and should NOT be included in the output.
- Explanations must be simple, easy to understand, and in plain text (not markdown).

**Output format:**
Plain text explanation in [TARGET_LANGUAGE], with a short supporting example in [NATIVE_LANGUAGE].

---

**Table 6: Prompt template for expression extraction, sentence translation, and expression mapping. The three stages are separated by horizontal rules. Placeholders such as [USER_INPUT], [NATIVE_LANGUAGE], [TARGET_LANGUAGE], [CONTEXT], and [TRANSLATED_TEXT], [LIST_OF_PHRASES] are dynamically replaced during runtime.**

---

**Stage 1: Extract Chinese Phrases**

**Instruction:**

You are a text analyzer.

The user is a non-native speaker of [TARGET_LANGUAGE];

His / Her native language is [NATIVE_LANGUAGE]

The user input may contain both [NATIVE_LANGUAGE] and [TARGET_LANGUAGE].

**Requirement:**

- Extract only meaningful [NATIVE_LANGUAGE] phrases that should be explained to a learner of [TARGET_LANGUAGE].
- Ignore URLs, numbers, emoji, and all non-Chinese characters.
- Deduplicate phrases; do not add explanations.

**Input:**

User input: [USER_INPUT]

**Output format:**

["短语1", "短语2"].

---

**Stage 2: Translate Full Sentence**

**Instruction:**

Translate the user's message (in [NATIVE_LANGUAGE]) into fluent [TARGET_LANGUAGE].

**Requirement:**

- Consider the given context only to understand the situation.
- Do not include or translate the context itself in the output.
- Ensure the translation is natural and fluent.

**Input:**

Context: [CONTEXT]

User Message: [USER_INPUT]

**Output format:**

{"translated_text": "..."}

---

**Stage 3: Map Chinese to English**

**Instruction:**

For each [NATIVE_LANGUAGE] phrase, find the corresponding exact [TARGET_LANGUAGE] phrase(s) in the translated sentence.

**Requirement:**

- Return results in the same order as the input Chinese phrases.
- Each item must be the exact phrase from the translated sentence.
- If a phrase is not found, return an empty string.

**Input:**

Original [NATIVE_LANGUAGE] phrases: [LIST_OF_PHRASES]

Translated sentence: "[TRANSLATED_TEXT]"

**Output format:**

["baby", "happy"]

**Table 7: Full instructions provided to native English speakers during the conversation task.**

| Section | Content |
|---|---|
| **Introduction** | Thank you for joining our study! Please take about 5 minutes to read the instructions below. |
| **Conversation Flow** | Please discuss the issues below with your conversation partner. You should lead the discussion with your conversation partner, as they are non-native English speakers and may not know how to raise these points. Begin with the first point. After 15 minutes, bring up the second point; after 30 minutes, bring up the third point. However, if the conversation is flowing naturally, feel free to ignore the time guidelines and continue discussing the current point. |
| **Important Note** | Treat your partner exactly as you would any other native English speaker. Do not simplify your grammar, vocabulary, sentence structures, or expressions. Use language as you normally would with friends, family, or colleagues. |
| **Conversation Style** | Chat as you would on WhatsApp or Telegram. Avoid long paragraphs; instead, send multiple shorter messages. |
| **Topic 1: Social Media and Youths** | Do youths today spend too much time on social media? Is growing up with social media positive or negative? How does this compare to people who did not grow up with constant social media? Does it affect development positively or negatively? |
| **Topic 2: Social Media as a Distraction** | Is modern social media a detriment to people in general? Do people consume too much "brainrot" or low-effort content? Does this reduce intrinsic cognitive effort or lead people to default to easy entertainment? Is this harmful? Does it detract from healthier or more productive activities, such as exercise or educational content? |
| **Topic 3: Social Media and Self-Image** | Modern social media emphasizes maintaining an idealized online self-image. Is this good or bad? Is there too much pressure to maintain a perfect image for others? Does this positively or negatively affect self-esteem? |